\documentstyle[preprint,aps]{revtex}
\def\lsim{\mathrel{\raise2pt\hbox to 8pt{\raise
-5pt\hbox{$\sim$}\hss{$<$}}}} \title{Pseudo-spin Symmetry and
Relativistic\\Single-nucleon Wave Functions}
\author{Joseph N.\ Ginocchio$^*$ and David G. Madland$^{\dagger}$}
\address{{\it Theoretical Division, Los Alamos National Laboratory\\ Los
Alamos, NM 87545\\ }}
\begin{document}
\draft
\maketitle
\begin{abstract}
\noindent We show that the occurrence of approximate pseudo-spin symmetry
in nuclei is
connected with certain similarities in the relativistic single-nucleon wave
functions
of the corresponding pseudo-spin doublets. We perform a case study in which
several examples and the systematics of this connection are explored.\\
 
\end{abstract}
\pagebreak
 
\section{INTRODUCTION}
 
In a recent paper \cite
{gino} it was shown that quasi-degenerate pseudo-spin doublets in nuclei,
discovered almost thirty years ago \cite{kth,aa}, arise from the near
equality in magnitude of
attractive scalar, $V_S$, and repulsive vector, $V_V$, relativistic mean
fields, $V_S \sim -V_V$, in which the nucleons move. Pseudo-spin doublets
have non-relativistic
quantum numbers ($n_r$, $\ell$,$j = \ell + 1/2)$ and ($n_{r}-1, \ell + 2$,
$j = \ell + 3/2$) where $n_r$,$\ell$, and $J$ are the single-nucleon
radial, orbital, and total
angular momentum quantum numbers, respectively. These authors defined a
``pseudo'' orbital angular momentum $\tilde{\ell}$ = $\ell$ + 1;
for example, $(n_r s_{1/2},(n_r-1) d_{3/2})$ will have $\tilde{\ell}= 1$ ,
$(n_r p_{3/2},(n_r-1)
f_{5/2})$ will have $\tilde{\ell}= 2$, etc. Then these doublets are almost
degenerate with
respect to ``pseudo'' spin, $\tilde s$ = 1/2, since $j = \tilde{\ell}\ \pm
\tilde s$ for the two states in the doublet. This symmetry has been used
to explain
a number of phenomena in nuclear structure \cite{bohr} including most
recently the identical rotational bands observed in nuclei \cite{ben},
and to establish an effective shell model coupling scheme \cite {draa}.
A near equality in the magnitude
of mean fields seems to be a universal feature of relativistic theories
ranging from relativistic field theories with interacting nucleons and
mesons
\cite {wal}, to nucleons interacting via Skyrme-type interactions
\cite{mad,madp},
to QCD sum rules \cite{furn}. In this paper we investigate the consequences
of $V_S\sim -V_V$ on the single-nucleon wave functions.
The connection between the Dirac equation and pseudo-spin symmetry is
reviewed in Sec. II. The relativistic mean field model described in Sec.
III is used for the
calculations of the case study presented in Sec. IV. Our conclusions are
given in Sec. V.
 
\section{THE DIRAC EQUATION AND PSEUDO-SPIN SYMMETRY}
 
The Dirac equation with external scalar, $V_S$, and vector, $V_V$,
potentials is given by:
\begin{equation}
[c{\bf {\alpha}}\cdot {{p}} + \beta (mc^2 + V_S) + V_V]\Psi = {\cal E}
\Psi,
\label {dirac}
\end{equation}
where ${\bf {\alpha}}$ and ${\bf{\beta}}$ are the usual Dirac matrices
\cite {mul}. We
shall be considering relativistic mean field theory with spherical symmetry
for which the scalar and vector
potentials depend only on the radial coordinate. In this case
the orbital angular momentum is not a conserved
quantum number in general. Instead a nucleon moving in a spherical
relativistic field is labeled by a radial quantum number, $n_r$, total
angular momentum $j$,
its projection on the z-axis, $m$, and ${\hat {\kappa}} = -{{\beta}}( {\hat
{\sigma}}\cdot{\hat {L}} + 1) $\cite {mul}.
The eigenvalues of
$\hat {\kappa}$ are
$\kappa =
\pm$(j + 1/2); -- for aligned spin ($s_{1/2},p_{3/2},$ etc.) and + for
unaligned spin ($p_{1/2},
d_{3/2}, $ etc.). Thus, the quantum number $\kappa$ and the radial quantum
number $n_{r}$ are sufficient to label the orbitals. The spherically
symmetric Dirac wave function can then
be written in terms of upper and lower components $\Psi_{\kappa > 0, m} =
(g_{\kappa} [Y_{\ell}\chi]_m^{(j = \ell - 1/2)}, if_{\kappa}[Y_{\ell -
1}\chi]_m^{(j = \ell -1/2)})$,
$\Psi_{\kappa < 0, m} = (g_{\kappa} [Y_{\ell}\chi]_m^{(j = \ell + 1/2)},
if_{\kappa}[Y_{\ell + 1}\chi]_m^{(j = \ell + 1/2)})$ where $g_{\kappa},
f_{\kappa}$ are the radial wave
functions (omitting the radial quantum numbers), $Y_{\ell}$ are the
spherical harmonics, $\chi$ is a two-component Pauli spinor, and
$[\dots]^{(j)}$
means coupled to angular momentum $j$.
The radial Dirac equation for the upper and lower components ($g_{\kappa},
f_{\kappa}$) of the single-nucleon radial wave function in dimensionless
units (also omitting the magnetic
quantum numbers) is then given by
\cite{mul}
\begin{equation}
\left[{d\over{dr}}+ {{ 1 + \kappa}\over r}\right] g_{\kappa} = \ [ 2 - E -
V(r) \
] f_{\kappa}
\label {d1}
\end{equation}
\begin{equation}
\left[{d\over{dr}}+ {{ 1 - \kappa}\over r}\right] f_{\kappa} = \ [E +
\Delta(r) \
]g_{\kappa},
\label {d2}
\end{equation}
where r is the radial coordinate in units of ${\hbar c}/{mc^2}$,
$V(r)$ = [$V_V(r)$ - $V_S(r)$] / $mc^2$, $\Delta(r)$ = [$V_S(r)$ +
$V_V(r)$] / $mc^2$, and $E$ is the
binding energy [$E = (mc^2 - {\cal E})/ mc^2> 0$] of the nucleon in
units of the free nucleon mass. First we repeat the proof \cite {gino}
that, in the limit of equality
of the  magnitude of the
vector and scalar potential, $\Delta( r )$ = 0, pseudo-spin is exactly
conserved. To do this, we solve for $g_{\kappa}$ in (\ref {d2}) and
substitute into (\ref{d1}), obtaining the
second order differential equation for $f_{\kappa}$, \begin{equation}
\left[{{d^2}\over{dx^2}}+ {2\over x}{{d}\over{dx}}-{{ \tilde{\ell} (
\tilde{\ell}
+ 1)}\over {x^2}}\ + ( V(r) - 2 + E )\right] f_{\kappa} = 0,
\label {s}
\end{equation}
where $x = \sqrt E\ r, E \ne 0$, and
 
\begin{equation}
\tilde{\ell} = \kappa - 1,\:\kappa > 0\:;\: \tilde{\ell} = -
\kappa,\:\kappa < 0,
\label {eq_k}
\end{equation}
which agrees with the original definition of the pseudo-orbital angular
momentum \cite {kth,aa}. For example,
for $(n_r s_{1/2},(n_r-1) d_{3/2})$, $\kappa = -1$ and $2$, respectively
giving $\tilde{\ell} =1$ in both cases. Furthermore, the physical
significance of $\tilde{\ell} $ is revealed; it is the ``orbital angular
momentum'' of the lower
component of the Dirac wave function and, in this limit, it is a conserved
quantum number.
 
For $ E \ne 0$, Equation (\ref {s}) is a Schr\"{o}dinger equation with an
attractive potential $V$ and binding energy $2 - E$ which depends only on
the pseudo-orbital angular
momentum,
$\tilde{\ell}$, through the pseudo - rotational kinetic energy, ${{
\tilde{\ell} ( \tilde{\ell}
+ 1)}\over {x^2}}$, and not on
$\kappa$. Hence the eigenenergies do not depend on $\kappa$ but only on $
\tilde{\ell}$. Thus the doublets with the same $ \tilde{\ell}$ but
different
$\kappa \ (\kappa = \tilde{\ell} + 1$ and $\kappa = - \tilde{\ell})$ will
be degenerate, producing pseudo-spin symmetry.
 
Furthermore, from (\ref{s}), it is clear that the $f_{\kappa}$ with the
same $ \tilde{\ell}$ can only differ by an\\ overall constant,
\begin{equation}
f_{-\tilde{\ell}} = {\cal N}_{\tilde{\ell}} \ f_{{\tilde{\ell}} +1 }.
\label {f}
\end{equation}
We show now that ${\cal N}_{\tilde{\ell}}^2 = 1$.
The Dirac wave function is normalized by
\begin{equation}
\int_0^{\infty} r^2 dr [ g_{\kappa}(r) ^2 + f_{\kappa}(r)^2 ] = 1. \label
{norm}
\end{equation}
Furthermore, from (\ref {d2}), we find in the pseudo-spin limit,
\begin{equation}
g_{{\tilde{\ell}} +1 }(r) = {1\over E_{\tilde{\ell}}} \left[{d\over dr } -
{{\tilde{\ell}}\over r}\right]f_{{\tilde{\ell}} +1}(r)\:,\:
g_{{-\tilde{\ell}} }(r) =
{1\over E_{\tilde{\ell}}} \left[{d\over dr } + {{\tilde{\ell}} + 1\over
r}\right]f_{-\tilde{\ell}} (r)\:, \label {gf}
\end{equation}
where we have used $E_{\tilde{\ell}} = E_{\tilde{\ell}+1} =
E_{-\tilde{\ell}}$. Using these relations in (\ref {norm}) for both
partners, $\kappa = - \tilde{\ell}$ and
$\kappa = \tilde{\ell} + 1$ and integrating by parts and then using the
differential equation (\ref {s}) for $f_{\kappa}$, we obtain
$$
\int_0^{\infty} r^2 dr f_{{\tilde{\ell}} +1}(r)^2 (2 - V - 2E_{\tilde{\ell}
}) = - E_{\tilde{\ell}},
$$
\begin{equation}
{\cal N}_{\tilde{\ell}}^2\int_0^{\infty} r^2 dr f_{{\tilde{\ell}} +1}(r)^2
(2 - V - 2E_{\tilde{\ell}}) = - E_{\tilde{\ell}},
\label {normeq1}
\end {equation}
which implies that ${\cal N}_{\tilde{\ell}}^2 = 1$ for $E_{\tilde{\ell}}
\ne 0 $. Furthermore, for a
square well \cite {gino}, ${\cal N}_{\tilde{\ell}} = -1$ and therefore we
assume that this is the case for more realistic nuclear mean fields
as well.
 
As pointed out in \cite {gino}, there will not be any bound Dirac valence
states, only Dirac sea states, for the pseudo-spin limit $\Delta = 0$,
which would contradict the fact that bound nuclei exist. In \cite {gino}
it was
shown that it is possible to have bound valence states $\it {and}$
quasi-degeneracy for a small $\Delta(r)$
by studying the Coulomb potential and the spherical potential well,
$V_{s\:,\:v}(r) = V_{s\:,\:v} > 0\:,\: r < R\:;\: V_{s\:,\:v}(r) = 0\:,\: r
> R$. Likewise, one may ask how well
the relation
\begin {equation} f_{-{\tilde{\ell}}}(r) = -f_{{\tilde{\ell}} +1}(r) \label
{frel}
\end {equation}
holds for small pseudo-spin breaking.
In this paper we employ a physically more realistic Dirac-Hartree approach
to answer this question.
 
\section{RELATIVISTIC POINT COUPLING MODEL}
 
We use a self-consistent Dirac-Hartree model with contact interactions
(point couplings) in the mean field ($\psi\;\longrightarrow\;\langle
\,\psi\,\rangle$) and no Dirac sea approximations. The model consists of
four-, six-, and eight-fermion point couplings leading to scalar and vector
densities with both
isoscalar and isovector components, derivatives of the densities to
simulate the finite ranges of the mesonic interactions,
but no explicit meson fields.
The Lagrangian is given by
 
\begin{equation}
{\cal L} = {\cal L}_{free} + {\cal L}_{4f} + {\cal L}_{hot} + {\cal
L}_{der} + {\cal L}_{em} \ , \ \ {\rm where}
\end{equation}
 
\vspace{6pt}
\noindent ${\cal L}_{free}$ and ${\cal L}_{em}$ are the kinetic and
electromagnetic terms, and
\begin{eqnarray}
{\cal L}_{4f} & = &
-\frac{1}{2}{{\alpha}_{S}}({\bar{\psi}}{\psi})({\bar{\psi}}{\psi})
-\frac{1}{2}{{\alpha}_{V}}({\bar{\psi}}{{\gamma}_{\mu}}{\psi})({\bar{\psi}}
{{\gamma}^{\mu}}{\psi}) \nonumber \\
& &
-\frac{1}{2}{{\alpha}_{TS}}({\bar{\psi}}{\vec{\tau}}{\psi}){\cdot}
({\bar{\psi}}{
\vec{\tau}}{\psi})
-\frac{1}{2}{{\alpha}_{TV}}({\bar{\psi}}{\vec{\tau}}{{\gamma}_{\mu}}{\psi})
{\cdot}
({\bar{\psi}}{\vec{\tau}}{{\gamma}^{\mu}}{\psi}) \ , \\
& & \nonumber \\
{\cal L}_{hot} & = & -\frac{1}{3}{{\beta}_{S}}({\bar{\psi}}{\psi})^{3}
-\frac{1}{4}{{\gamma}_{S}}({\bar{\psi}}{\psi})^{4} \nonumber \\ & &
-\frac{1}{4}{{\gamma}_{V}}[({\bar{\psi}}{{\gamma}_{\mu}}{\psi})
({\bar{\psi}}
{{\gamma}^{\mu}}
{\psi})]^{2} \ , \ {\rm and} \\
& & \nonumber \\
{\cal L}_{der} & = &
-\frac{1}{2}{{\delta}_{S}}({\partial_{\nu}}{\bar{\psi}}\psi)
({\partial^{\nu}}{\bar{\psi}}\psi) \nonumber \\ & &
-\frac{1}{2}{{\delta}_{V}}({\partial_{\nu}}{\bar{\psi}}
{\gamma_{\mu}}\psi) ({\partial^{\nu}}{\bar{\psi}}{\gamma^{\mu}}\psi) \ .
\end{eqnarray}
 
In these equations, $\psi$ is the nucleon field, the subscripts ``$S$'' and
``$V$'' refer to the scalar and vector nucleon fields, respectively, and
the subscript ``$T$''
refers to isovector fields containing the nucleon isospin ${\vec{\tau}}$.
The physical makeup of ${\cal L}$ is that ${\cal L}_{4f}$ is a four-fermion
interaction, while ${\cal
L}_{hot}$ contains higher order six-fermion and eight-fermion interactions,
and ${\cal
L}_{der}$ contains derivatives in the nucleon densities. There are a total
of nine coupling constants.
 
Minimizing the expectation value of the Hamiltonian corresponding to Eq.
(11) in the space of Slater determinants $|\phi\rangle$ leads to the
Dirac-Hartree
equations containing the following potentials: \begin{eqnarray}
{V_{S}} & = & {\alpha_{S}}{\rho_{S}} + {\beta_{S}}{\rho_{S}^{2}} +
{\gamma_{S}}{\rho_
{S}^{3}} + {\delta_{S}}{\Delta}{\rho_{S}} \ , \\ {V_{V}} & = &
{\alpha_{V}}{\rho_{V}} + {\gamma_{V}}{\rho_{V}^{3}} + {\delta_{V}}
{\Delta}{\rho_{V}} \ , \\
{V_{TS}} & = & {\alpha_{TS}}{\rho_{TS}} \ , \ {\rm and} \\ {V_{TV}} & = &
{\alpha_{TV}}{\rho_{TV}} \ , \end{eqnarray}
 
\noindent where Eq. (15) is the isoscalar-scalar potential corresponding to
$\sigma$ meson (fictitious) exchange, Eq. (16) is the isoscalar-vector
potential corresponding to
$\omega$ meson exchange, Eq. (17) is the isovector-scalar potential
corresponding to $\delta$ meson exchange, and Eq. (18) is the
isovector-vector potential corresponding to $\rho$
meson exchange. In these latter equations the scalar density is given by
$\rho_{S}\;=\;\langle \phi|{\bar{\psi}}{\psi}|\phi\rangle$, the vector
density is given by $\rho_{V}\;
=\;\langle\phi|{\bar{\psi}}{\gamma_{0}}{\psi}|\phi\rangle$, the
isovector-scalar density is given by
$\rho_{TS}\;=\;\langle\phi|{\bar{\psi}} {\tau_{3}}{\psi}|\phi\rangle$, and
the
isovector-vector density is given by $\rho_{TV}\;=\;\langle\phi|
{\bar{\psi}}{\tau_{3}}{\gamma_{0}}{\psi}|\phi\rangle$. These densities can
of course be rewritten in terms of the
upper and lower components of the Dirac wave functions
($g_{\alpha},f_{\alpha}$), respectively, where $\alpha = \{n_{r},\kappa
,m\}$.
For example, the vector (baryon)
density becomes \begin{equation}
\rho_{V}(r) = \sum_{\alpha}^{occ.}\frac{2j + 1}{4\pi}[g_{\alpha}^{2}(r) +
f_{\alpha}^{2}(r)]\;,
\end{equation}
while the scalar density becomes
\begin{equation}
\rho_{S}(r) = \sum_{\alpha}^{occ.}\frac{2j + 1}{4\pi}[g_{\alpha}^{2}(r) -
f_{\alpha}^{2}(r)]\;.
\end{equation}
 
The nine coupling constants of the Lagrangian were determined in a
self-consistent procedure that solves the Dirac-Hartree equations for
several nuclei simultaneously in a nonlinear
least-squares adjustment algorithm of Levenberg-Marquardt type with respect
to well-measured nuclear ground-state observables.
The well-measured observables used (in order of decreasing experimental
accuracy) are (a) the
ground-state masses (binding energies),
(b) the rms charge radii, and (c) the spin-orbit splittings of the
least-bound neutron and proton spin-orbit pairs. We chose these observables
for ten spherical closed-shell nuclei
ranging from $^{16}$O to $^{208}$Pb in the determination of the coupling
constants (40 observables to determine 9 coupling constants). The final
coupling constants \cite{madp} are
given in Table I where the first four refer to Eq. (12), the next three
refer to Eq. (13), and the remaining two refer to Eq. (14).
They span 13 orders of magnitude, but seven of them and the sum of the
remaining two are {\it natural} (dimensionless numbers of order 1) when
scaled in accordance with QCD
mass scales and taking into account the constraint of chiral symmetry
\cite{madp,madpp}. They should be viewed as an interim set of coupling
constants in the development of the
relativistic point coupling model. Note that the initial set of coupling
constants determined with this approach appears in Ref. \cite{mad}.
 
With these nine coupling constants one can calculate the following for
spherical closed-shell nuclei: (a) Dirac single-nucleon wave functions and
eigenvalues for both protons
and neutrons, (b) nuclear ground-state mass and binding energy, (c) proton
and neutron densities and their moments, (d) nuclear charge density and its
moments, and (e) isoscalar- and
isovector-, scalar and vector, potentials.
 
\section{CASE STUDY}
 
For our present purpose we consider the upper and lower components of the
Dirac single-nucleon wave functions for $^{208}$Pb. Using the coupling
constants of Table I
we have calculated the single-neutron wave functions for the occupied
orbits in the ground state of this nucleus; these orbits have six neutron
pseudo-spin
doublets. Their properties are listed in Table II. Similarly, we have
calculated the single-proton wave functions for the occupied orbits in
the ground state of the same nucleus
which contain four proton pseudo-spin doublets, and their properties are
listed in Table III. The pseudo-spin doublets are ordered in the two tables
according to
increasing orbital angular momentum
$\tilde{\ell}$ of the lower Dirac component
$f_{n_{r},\kappa}$ , for each value of the radial quantum number $n_{r}$.
The energy eigenvalues therein, $\varepsilon_{n_{r},\kappa}$ , in units of
MeV, are related to the
dimensionless energy eigenvalues
$E_{n_{r},\kappa}$, by
\begin{equation}
\varepsilon_{n_{r},\kappa} = E_{n_{r},\kappa}\:mc^{2} \end{equation}
where $mc^{2}$ is either the free neutron mass or the free proton mass.
 
The upper and lower components of the Dirac wave functions for the most
deeply bound neutron and proton pseudo-spin doublets (first entries in
Tables II and III,
respectively) are shown in Figs. 1 -- 4. In Figs. 2 and 4 the lower
components of the doublets
are plotted with opposite sign so as to directly test Eq. (10), the
pseudo-spin limit, against the observed (calculated) pseudo-spin breaking.
It is clear that Eq. (10) is crudely
satisfied when the eigenvalues are crudely degenerate (to within $\sim$ 3
MeV for both neutrons and protons in this example). Comparisons of Figs. 1
and 2, for neutrons, and Figs. 3
and 4, for protons, show that while the upper components of the total Dirac
wave function dominate, the lower components are certainly not negligible,
and approximate, although crude
in this example, pseudo-spin symmetry exists for both protons and neutrons.
 
We examine the systematics of pseudo-spin symmetry and the corresponding
relativistic wave functions by considering the six neutron pseudo-spin
doublets in Table II.
Figures 5 -- 7 show the lower components of three of the remaining five
doublets. In
Figs. 5 and 6, for $\tilde{\ell}$ = 2 and 4, respectively, the pseudo-spin
limit given by Eq. (10) is
again approximately satisfied to
about the same degree as that for the first neutron pseudo-spin doublet
for $\tilde{\ell}$ = 1 shown in Fig. 2.
The same remark applies to the case for $\tilde{\ell}$ = 3 (not shown).
The four doublets all have the same radial quantum number $n_{r}
= 1$, and have increasing orbital
angular momentum $\tilde{\ell}$ of the lower component. Figure 7
shows the lower components for the first of the two neutron pseudo-spin
doublets
that have radial quantum number
$n_{r} = 2$. Here there is dramatic improvement in pseudo-spin symmetry, as
expected \cite{gino} because these states are less bound and exact
pseudo-spin symmetry occurs for {\it no
bound} Dirac valence states. The same is true for the second neutron
pseudo-spin doublet with $n_{r} = 2$ in Table II (not shown).
A related systematic is that if the Dirac
single-nucleon states of the pseudo-spin doublets of Tables II and III are
sorted by classical major oscillator
shells, then for those states in the same major shell, the energy splitting
of the doublets decreases as $\tilde{\ell}$ decreases. In Table II, the
($n_{r},\tilde{\ell}$) = (1,3) and
(2,1) doublets are in the $N$ = 4 major shell and the (1,4) and (2,2)
doublets are in the $N$ = 5 major shell. In both cases the energy splitting
decreases with decreasing
$\tilde{\ell}$. The same is true for the (1,3) and (2,1) doublets in Table
III also corresponding to the $N$ = 4 major shell.
This point becomes more clear in Fig. 8 which shows the spectra for the
twelve states making up the six neutron doublets and the eight states
making up the four proton doublets.
Clearly, for the doublets with the same pseudo-orbital
angular momentum, the splitting between doublets
decreases as the binding energy decreases, or, equivalently, as the radial
quantum number increases. Also
for doublets with different pseudo-orbital angular momentum, but roughly
the same binding energy, say for example, ($n_{r},\tilde{\ell}$) = (1,3)
and (2,1), and ($n_{r},\tilde{\ell}$) = (1,4) and (2,2), the doublet
with the
lower ${\tilde \ell}$ has the smaller energy splitting.  These features
are seen in the square well
potential as in Equation (19) in \cite {gino} in which the energy splitting
between the doublets is proportional to $2{\tilde \ell} + 1 $ and $ E$.
The six neutron pseudo-spin doublet
splittings of Fig. 8 are again plotted in Fig. 9, with the factor
(2$\tilde{\ell}$ + 1) divided out, {\it vs} the mean eigenvalue
$<\:\varepsilon \:>$ for each doublet. Here it is clearly quantitative that
the energy splitting decreases as the binding energy decreases.
 
Finally, we compare two calculated neutron pseudo-spin doublets with
experiment in Table II and two calculated proton pseudo-spin doublets with
experiment in Table III.
For both protons and neutrons, the magnitudes of the calculated eigenvalues
and corresponding measured values are comparable and the calculated energy
splittings agree to within
a factor $\sim$2 with the measured energy splittings in each of the four
cases. The calculated splittings are always larger than the measured
splittings which is not very surprising
as most bound-state single-nucleon
eigenvalue calculations, in relativistic mean field approaches, share the
common feature of ``spread-out eigenvalues'' (see, for example, Ref.
\cite{wal,mad}) and our interim coupling constants (Table I) provide no
exception. In addition, the measured and calculated states making up the
neutron ($n_{r},\tilde{\ell}$) = (2,2) pseudo-spin doublet in Table II are
reversed, a feature also occurring in other relativistic calculations \cite
{wal,mad}.
On the other hand, the measured and calculated states making up the neutron
(1,4) pseudo-spin doublet in the same table are not reversed in our
calculation, but are reversed in
other recent non-relativistic calculations \cite {ndref,mn1}.
We remark here that these latter features do not change our observations as
to the connections between (approximate) pseudo-spin symmetry and the
corresponding Dirac single-nucleon
wave functions and eigenvalues.
 
\section{CONCLUSIONS}
 
We have shown that physically realistic relativistic mean fields lead to
small pseudo-spin symmetry breaking which implies that the lower components
of the corresponding Dirac single-nucleon wave functions satisfy
\begin {equation}
f_{-{\tilde{\ell}}}(r) \simeq -f_{{\tilde{\ell} } +1}(r)
\label {lower}
\end {equation}
where the pseudo-orbital angular momentum $\tilde{\ell}$ is
exactly conserved in the pseudo-spin limit.
Near this limit the lower components depend only upon $\tilde{\ell}$
(for fixed radial quantum number $n_{r}$) and the energy splitting
between the doublets is small. Furthermore the pseudo-spin symmetry
becomes increasingly valid as the pseudo-orbital angular momentum
$\tilde{\ell}$ decreases and as the binding energy $E_{\tilde{\ell}}$
decreases (or as the radial quantum number $n_{r}$ increases). These
observations and conclusions are based upon our calculations of ten
pseudo-spin doublets occurring in $^{208}$Pb of which four have been
measured and compare reasonably well with the corresponding calculated
doublets. Thus this work confirms the contention of Ref. \cite{gino}
that pseudo-spin symmetry in nuclei arises largely from nucleons moving in
relativistic mean fields with attractive isoscalar-scalar and repulsive
isoscalar-vector components that are nearly equal in magnitude.
 
\section { Acknowledgments}
 
This research was supported by the United States Department of
Energy.\\
 
\noindent *Electronic address: gino@t5.lanl.gov\\
$^{\dagger}$Electronic address: dgm@lanl.gov
\pagebreak

\begin{table}
\caption{Optimized Coupling Constants for the Interim Relativistic Point
Coupling Model}
\begin{tabular}{ccc} Coupling Constant & Magnitude & Dimension \\
\tableline
$\alpha_{S}$ & -4.517${\times}10^{-4}$ & MeV$^{-2}$ \\ $\alpha_{TS}$ &
-2.168${\times}10^{-5}$ & MeV$^{-2}$ \\ $\alpha_{V}$ &
3.435${\times}10^{-4}$ & MeV$^{-2}$
\\ $\alpha_{TV}$ & 5.365${\times}10^{-5}$ & MeV$^{-2}$ \\ $\beta_{S}$ &
1.137${\times}10^{-11}$ & MeV$^{-5}$ \\ $\gamma_{S}$ &
5.731${\times}10^{-17}$ & MeV$^{-8}$ \\ $\gamma_{V}$ &
-4.423${\times}10^{-17}$ & MeV$^{-8}$ \\ $\delta_{S}$ &
-4.282${\times}10^{-10}$ & MeV$^{-4}$ \\ $\delta_{V}$ &
-1.155${\times}10^{-10}$ & MeV$^{-4}$ \\
\end{tabular}
\end{table}
 
\begin{table}
\caption{Calculated and Measured Neutron Pseudo-Spin Doublets in
$^{208}$Pb}
\begin{tabular}{ccccccc}
$\tilde{\ell}$ & $n_{r},\kappa{< 0}$ & ($\ell,j$) &
$\varepsilon_{n_{r},\kappa{< 0}}$
& $n_{r}-1,\kappa{> 0}$ & ($\ell+2,j+1$) & $\varepsilon_{n_{r}-1,\kappa{>
0}}$ \\
& & & (MeV) & & & (MeV) \\
\tableline
1 & 1, -1 & ($s_{1/2}$) & 42.139 & 0, 2 & ($d_{3/2}$) & 45.331 \\ 2 & 1, -2
& ($p_{3/2}$) & 31.454 & 0, 3 & ($f_{5/2}$) & 35.490 \\ 3 & 1, -3 &
($d_{5/2}$) & 20.981 & 0, 4 & ($g_{7/2}$) & 24.741 \\ 4 & 1, -4 &
($f_{7/2}$) & 10.944 & 0, 5 & ($h_{9/2}$) & 13.519 \\ 4 & 1, -4 &
($f_{7/2}$) & 9.708\tablenote{Experimental value from Ref.\ \cite{aaa}.} &
0, 5 & ($h_{9/2}$) & 10.781$^{\rm a}$ \\
\tableline
1 & 2, -1 & ($s_{1/2}$) & 18.129 & 1, 2 & ($d_{3/2}$) & 19.002 \\ 2 & 2, -2
& ($p_{3/2}$) & 7.656 & 1, 3 & ($f_{5/2}$) & 8.353 \\ 2 & 2, -2 &
($p_{3/2}$) & 8.266$^{\rm a}$ & 1, 3 & ($f_{5/2}$) & 7.938$^{\rm a}$ \\
\end{tabular}
\end{table}
 
\begin{table}
\caption{Calculated and Measured Proton Pseudo-Spin Doublets in $^{208}$Pb}
\begin{tabular}{ccccccc}
$\tilde{\ell}$ & $n_{r},\kappa{< 0}$ & ($\ell,j$) &
$\varepsilon_{n_{r},\kappa{< 0}}$
& $n_{r}-1,\kappa{> 0}$ & ($\ell+2,j+1$) & $\varepsilon_{n_{r}-1,\kappa{>
0}}$ \\
& & & (MeV) & & & (MeV) \\
\tableline
1 & 1, -1 & ($s_{1/2}$) & 32.047 & 0, 2 & ($d_{3/2}$) & 35.830 \\ 2 & 1, -2
& ($p_{3/2}$) & 21.800 & 0, 3 & ($f_{5/2}$) & 26.358 \\ 3 & 1, -3 &
($d_{5/2}$) & 11.597 & 0, 4 & ($g_{7/2}$) & 15.930 \\ 3 & 1, -3 &
($d_{5/2}$) & 9.696\tablenote{Experimental value from Ref.\ \cite{aaa}.} &
0, 4 & ($g_{7/2}$) & 11.487$^{\rm a}$ \\
\tableline
1 & 2, -1 & ($s_{1/2}$) & 8.416 & 1, 2 & ($d_{3/2}$) & 9.663 \\ 1 & 2, -1 &
($s_{1/2}$) & 8.013$^{\rm a}$ & 1, 2 & ($d_{3/2}$) & 8.364$^{\rm a}$
\end{tabular}
\end{table}
 
\begin{figure}
\caption{Dirac upper component wave functions for the neutron pseudo-spin
doublet ($n_{r},\tilde{\ell}$) = (1,1) in $^{208}$Pb and specified more
completely in the first line of Table II.}
\label{1}
\end{figure}
 
\begin{figure}
\caption{Dirac lower component wave functions corresponding to Fig. 1. The
sign of the component with $\kappa > 0$ has been reversed to test Eq.
(10).}
\label{2}
\end{figure}
 
\begin{figure}
\caption{Dirac upper component wave functions for the proton pseudo-spin
doublet ($n_{r},\tilde{\ell}$) = (1,1) in $^{208}$Pb and specified more
completely in the first line of Table III.}
\label{3}
\end{figure}
 
\begin{figure}
\caption{Dirac lower component wave functions corresponding to Fig. 3. The
sign of the component with $\kappa > 0$ has been reversed to test Eq.
(10).}
\label{4}
\end{figure}
 
\begin{figure}
\caption{Dirac lower component wave functions for the neutron pseudo-spin
doublet ($n_{r},\tilde{\ell}$) = (1,2) in $^{208}$Pb and specified more
completely in the second line of Table II. The sign of the component with
$\kappa > 0$ has been reversed to test Eq. (10).}
\label{5}
\end{figure}
 
\begin{figure}
\caption{The same as Fig. 5 except for the pseudo-spin doublet
($n_{r},\tilde{\ell}$) = (1,4) and specified more completely in the fourth
line of Table II.}
\label{6}
\end{figure}
 
\begin{figure}
\caption{The same as Fig. 5 except for the pseudo-spin doublet
($n_{r},\tilde{\ell}$) = (2,1) and specified more completely in the sixth
line of Table II.}
\label{7}
\end{figure}
 
\begin{figure}
\caption{Spectra of the six neutron and four proton pseudo-spin
doublets occurring in $^{208}$Pb. Note that the pseudo-spin symmetry
is more valid for the higher value of the radial quantum number.}
\label{8}
\end{figure}
 
\begin{figure}
\caption{Dirac energy splittings of the six neutron pseudo-spin doublets
shown in Fig. 8, normalized by (2$\tilde{\ell}$ + 1), {\it vs} the mean
energy eigenvalue $< \varepsilon >$ of the doublet.
The line is a linear fit to the four cases with $n_{r}$ = 1.}
\label{9}
\end{figure}
 
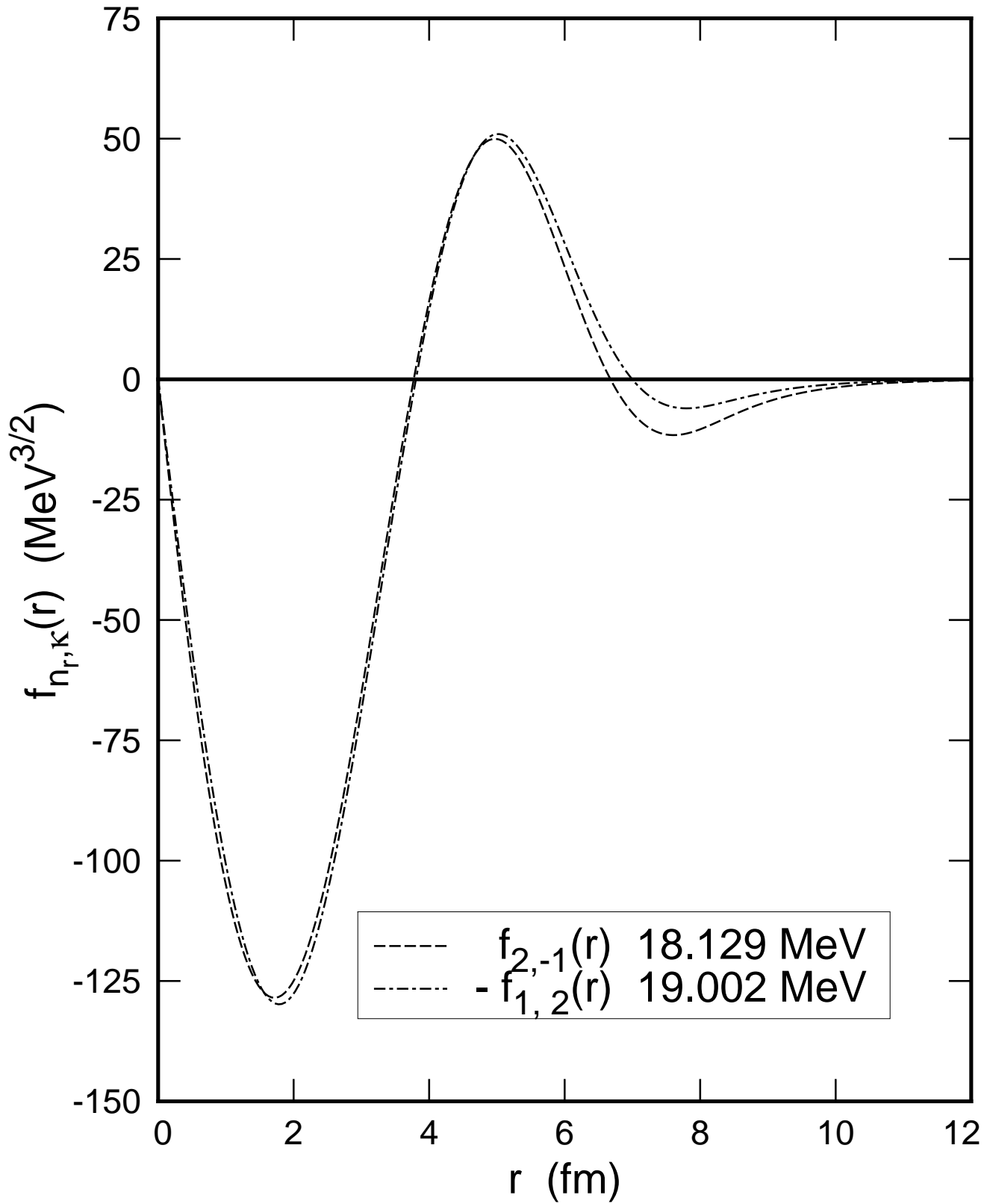
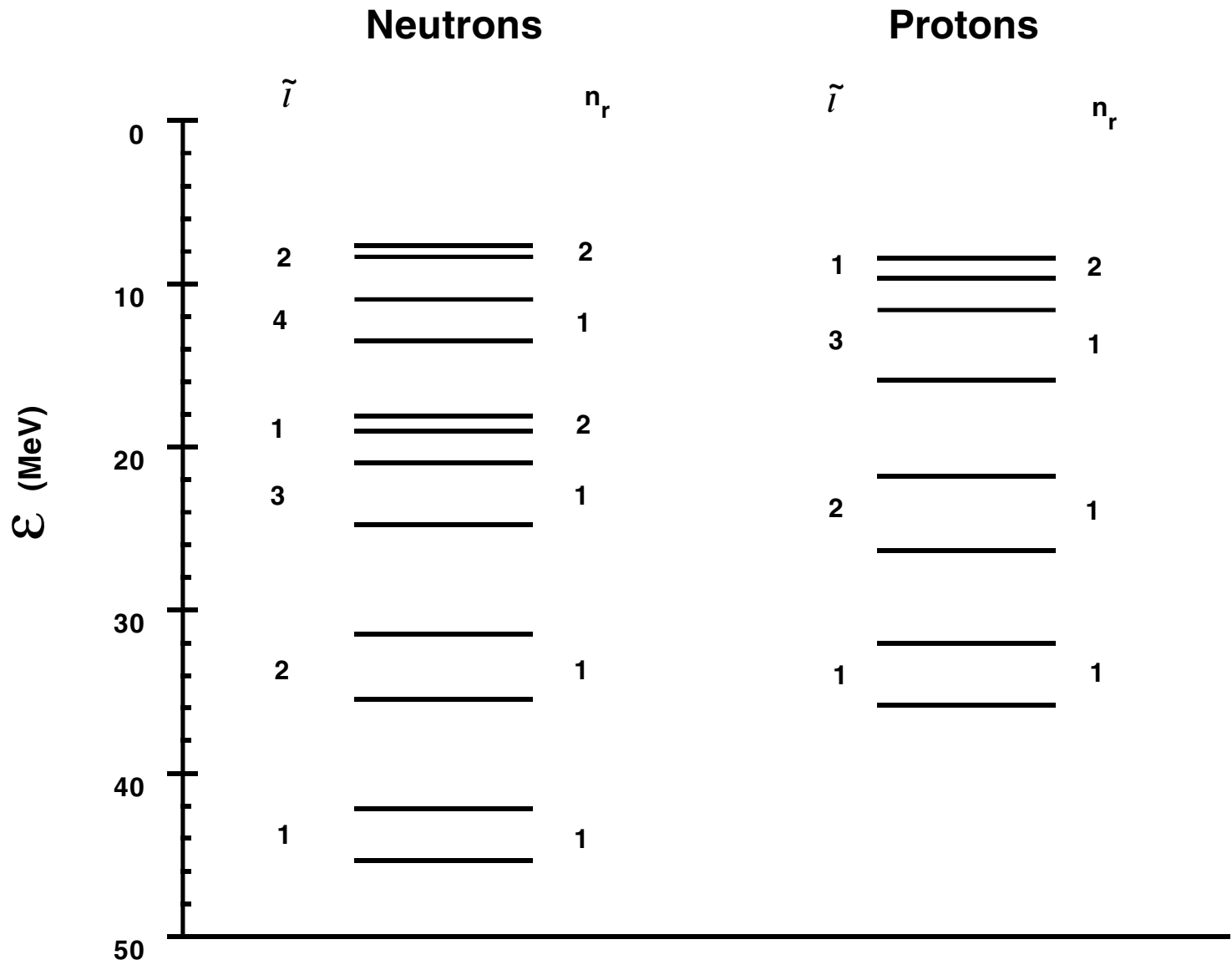
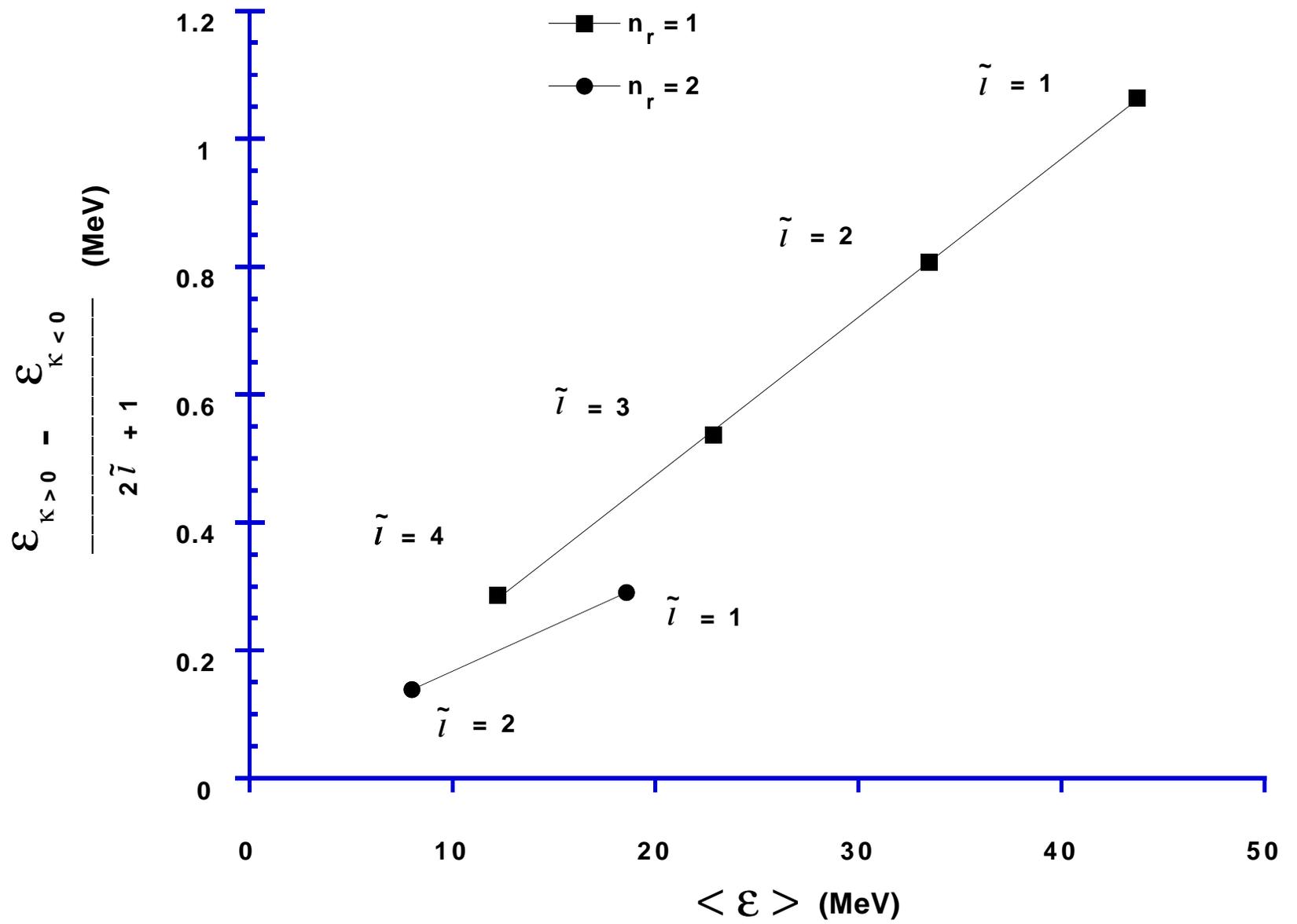
\end{document}